# Combined digital drone camera and optical channel parameters for air surveillance

Wamidh Jalil Mazher[1], hadeel Tariq Ibrahim[2],

[1]Department of Electrical Engineering, Southern TechnicalUniversity, Basra, Iraq.
[2]E-leraning Department, Thi qar University, Thi qar, Iraq
E-mail: wamidh.mazher@stu.edu.iq[1] ,Hadeel.Tariq@utq.edu.iq[2]

*Abstract*-Digital drone cameras with free-space optical (FSO) communication networks have been proposed to be promising for air surveillance. In the FSO channel, atmospheric turbulence (AT) degrades the signal. In this study, to mitigate the AT effect, we combined the parameters of both the digital drone camera and the optical channel. To support this proposal, the digital drone camera parameters are indicated by the field of view and camera–object distance. Meanwhile, the optical channel parameters, rather than the altitude, are denoted by the most critical parameter, which is the refractive index structure parameter used to characterize the effects of AT. Consequently, two lemmas are proposed and combined to present the optimum relationship between the digital drone camera and optical channel parameters. Therefore, the quality of the entire air surveillance system with a digital drone camera-FSO is significantly improved. Furthermore, the analysis and optimization for practical cases were applied to support our findings. Finally, our results demonstrated that an impressive performance improvement of an air surveillance system of 17 dB is possible compared to without optimization by combining digital drone camera and FSO parameters at a target outage transceiver probability of $10^{-6}$.

*Index Terms*- camera parameters, optical channel, AT, optimization, air surveillance.

## 1. INTRODUCTION

Drones have been widely used in several air surveillance applications [1], [2] A digital drone camera captures images of a phenomenon. The images are thereafter sent as an on-air video bit stream to a monitoring station (MS)[2]. In [2], the signals from the drone were sent to the MS by an optical wireless communication system. In [3], for a general digital drone camera swarm with an optical communication system for a deep decision, a suspending problem was configured and investigated, whereas in [4], to send the image using a free-space optical (FSO) system, the optical transceiver was considered, where each drone converted the digital camera image from the electric to optical domain using an onboard electrical/optical (E/O) convertor system and used a common modulator with direct detection intensity modulation[5].

The FSO communication system is intended for a drone communicating with an MS [6], [7] in air surveillance. This utility is due to the facilitation of higher throughput of the FSO system, which is in terabits per second per Hertz (Tb/s/Hz). Despite the major advantages of FSO systems, atmospheric turbulence (AT)-induced signal fading is a major factor that degrades the performance of optical signals[8]. Therefore, extra optical power marginal (*PM*) of an optical signal due to AT is needed.[9] The degradation has a nonlinear influence with respect to optical channel altitude (𝔸).[10] Accordingly, owing to passing through the optical channel, the drone signals are degraded, resulting in the fading of the entire air surveillance system performance. The major techniques utilized to mitigate this degradation are relay assistance[11]–[13] and utilization of long codewords for optical decoding/ forwarding[14]. The drawback of relay assistance and utilization of long codewords for optical decoding/forwarding as mitigating techniques is that they do not work with individual drone-FSO systems. In contrast, our proposal is applicable to a single drone. In this study, the optical signal degradation was reduced by combining the two major air surveillance subsystems.

The air surveillance system can be divided into two major subsystems; the digital drone camera and the communication system that utilizes a signal transceiver between the drone and MS. Therefore, the camera and optical subsystem parameters were correlated, and combining and optimizing parameters across both subsystems in advance led to an increase in surveillance system efficiency.

In this study, the digital drone camera field of view (*FOV*) and the drone-camera–object distance will be optimized and combined with the optical channel altitude. This proposal is a solution for optical channel degradation by AT and a mitigation strategy. New lemmas are presented to describe the relationship between the *FOV* and *PM* of a digital drone camera with an optical channel (i.e., combined).

The remainder of this paper is organized as follows. Section 2, presents the system modeling with subsections for optical channel parameters with the influence of AT on them and digital drone camera parameters performance that related to the optical channel. Section 3 presents the optimized and combined optical channel and digital camera parameters of the network drone. Results and discussion of air surveillance performance are presented in Section 4 before finally providing the study's conclusion in Section 5.

2. System Modeling

In our proposed system, a video bit stream signal is generated from a digital drone camera ($D_i$) and thereafter sent to an MS by an optical channel connection, as shown in Figure 1.

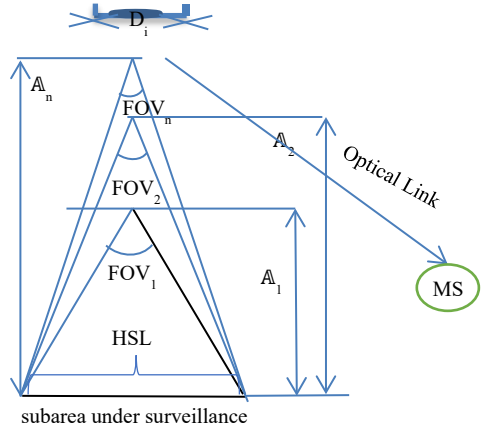

**Fig. 1.** Illustration of *FOV* and $\mathbb{A}$ for the digital drone camera-based FSO network for a preselected area under air surveillance

Our system for air surveillance has two major subsystems: the optical subsystem and the digital drone camera subsystem. Both the two subsystems' parameters of the entire air surveillance system are associated by a key parameter, which is altitude ($\mathbb{A}$), where it is same for the optical channel and camera as shown in Fig. 1.

Because the proposal to mitigate the signal degradation in the optical domain is proposed by combining these two parts, the system modeling section will be discussed in two subsections: optical channel and digital camera subsections.

2.1 Optical Channel

As mentioned, the optical signal is degraded by AT while passing through the optical channel, and a *PM* is required. Thus, the outage probability (Po) of the optical link (*L*) failure is equal to or less than the Poth value, where Poth denotes the outage probability threshold of the optical channel in the absence of AT effect. Accordingly, the *PM* required owing to AT degradation is determined by[9] [(Eq. 6.43), P.P. 378] as follows:

$$P_M = exp\left(\sqrt{-2\sigma C_n^{2^2} ln2p_0 + \sigma C_n^{2^2}/2}\right) \qquad (1)$$

where $\sigma = k\left(\frac{2\pi}{\lambda}\right)^{7/6} L^{11/6}$ is a preselected constant parameter[15] of the optical communication system (see Table 2 in the Results section), k is the optical wave number obtained from wavelength (λ) [10], and L is the optical channel length. In (1), the scintillation index ($C_n^2$) is the major key to the AT effect, and its optical channel altitude (𝔸) is inversely proportional to the contingent. $C_n^2$ is obtained as follows:[10]

$$c_n^2(\mathbb{A}) = 3.6 \times 10^{-3}(10^{-5}\mathbb{A})^{10} exp\left(-\frac{\mathbb{A}}{1000}\right) + 27 \times 10^{-16} exp\left(-\frac{\mathbb{A}}{1500}\right) + C_n^2(0) exp\left(-\frac{\mathbb{A}}{100}\right) \qquad (2)$$

where $C_n^2(0)$ denotes the refractive index structure parameter value on the ground in m-2/3 units.

Clearly, in (2), the value of 𝔸 is optimized, the scintillation index value is minimized, and *PM* is reduced according to Eq. 1. Consequently, the optical system performance increases, and the quality of the air surveillance system by the digital drone camera-optical channel also increases. To optimize the value of 𝔸, and because 𝔸 is a digital drone, camera parameters are related; therefore, a digital drone camera parameter description and analysis is needed. This is provided in the next subsection.

*2.2 Digital Drone Camera*

For a specified camera, the image (or video) captures the quality impacts by two major factors: the camera *FOV* and the camera–object distance (*COD*). It is easy to show the relationship between them (that is, *FOV* and *COD*) from the geometry in Fig. 2 and given by Eq. 3 as follows:

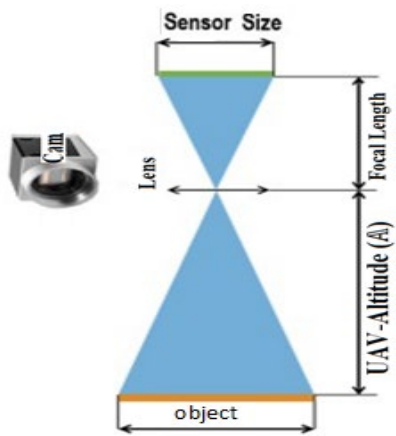

Fig. 2. Illustration of the camera objectives with UAV (drone) altitude of the area under air surveillance

$$\mathbb{A} = c_1 \cdot (tan(FOV/2))^{-1} \tag{3}$$

where $c_1 = \frac{1}{2}\frac{C_p}{Ï}$ denotes a constant value referring to the captured image requirements[2] that are image resolution ($Ï$) in Pix/m, as listed in Table 1, and $C_p$, which is the camera horizontal pixel number in Pix.

The *FOV* value depends on the camera focal length, where the *COD* and both the image resolution and number of camera-pixel values are related. In addition, the generality of this study is not lost, and both the camera focal length and camera-pixel availability are assumed. The image resolution based on the applications was classified, as summarized in Table 1.

Table-1 - Image resolution roughly classified[2]

| Application | Observation & detection | Recognition | Identification | Unit |
|---|---|---|---|---|
| Pix No. | range ≤ 30 | 30–120 | 120–150 | pixels/ft |

The drone *COD* value is equal to the optical channel altitude ($\mathbb{A}$), and this is clear in Fig. 2; therefore,

$$COD = \mathbb{A} \tag{4}$$

To optimize altitude $\mathbb{A}$ of the entire air surveillance system, both the digital drone camera and the optical-channel parameters are combined. Thereafter, the AT optical channel is decreased according to Eq. 2. Consequently, from Eq. 1, the *PM* value also decreases such that from this optimization, the entire air surveillance system quality is augmented. Therefore, this combination of digital drone cameras with optical channel parameters will be obtained in the next section.

Combined optical channel and digital camera parameters of network drone

The purpose of combining both optical channels with digital drone camera parameters is to mitigate the atmospheric-turbulence induced, and thereafter improve the entire air surveillance system. This will be achieved by optimizing the digital drone camera parameters related to the optical channel parameters, as shown in the optimization flowchart in Fig. 3.

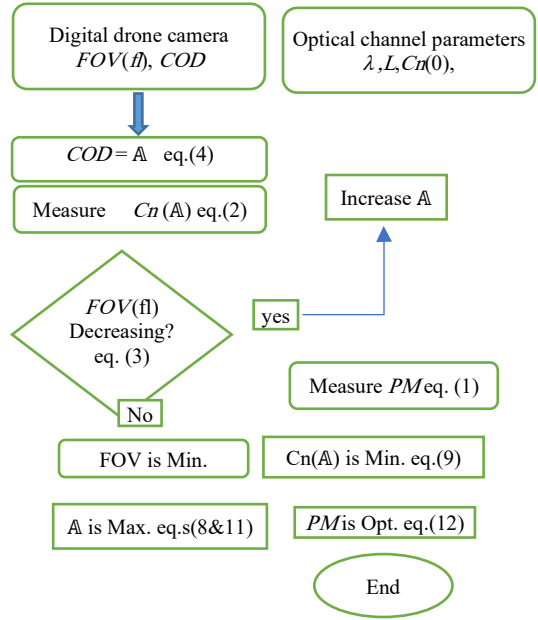

Fig. 3. Optimized flowchart of digital drone camera-optical channel parameters for air surveillance.

To support the drone parameters combination (digital camera–optical channel), a quick mathematical modification of Eq. 2 was prepared. The rewriting process is as follows:

$$c_n^2(\mathbb{A}) = \delta_1 \mathbb{A} e^{-\mathbb{A}} \qquad (5)$$

where $\delta_1$ denotes a constant related to the preselected parameters in Eq. 2. From Eq. 1, we obtain the following identity:

$$P_M \equiv \delta_2 e^{-\mathbb{A}} \qquad (6)$$

where $\delta_2$ denotes another constant related to the constant parameters of Eq. 1. From Eq. 1 & 6, we can easily show that:

$$\mathbb{A}_{opt} = \mathbb{A}_{max}, \qquad (7)$$

where $\mathbb{A}_{opt}$ denotes the optimum value of $\mathbb{A}$, which is a satisfactory $P_M$ value, and it is the minimum (optimum) according to Eq. 6. Finally, because $\mathbb{A}_{max}$ in Eq. 7 is related to the digital drone camera parameters mentioned previously, Eq. 3 can be simply expressed by the following:

$$\mathbb{A}_{max} = c_1 \cdot (tan(\ FOV_{min}/2))^{-1} \qquad (8)$$

**Lemma 1**: Assuming that $FOV = \{FOV_{min.}, FOV_{max.}\}$ is an available set; then, the maximum value of $\mathbb{A}$ is the optimum ($\mathbb{A}_{opt} = \mathbb{A}_{max}$) at the minimum $FOV$ value. Therefore,

$$C^2_{n_{i_{min.}}} \equiv FOV_{i_{min.}}, \text{ and i is unique} \tag{9}$$

Proof: If $FOV$ is not minimum at i; then, assume that:

$$C^2_{n_{i_{min}}} \equiv FOV_{q_{min}}, \tag{10}$$

where $q \neq i$. By setting $FOV = AOVj$, where $AOVj$ is the angle of view for the captured image (j), such that the image quality that is classified as listed in Table 1 is satisfied, we obtain the following from Eqs. 7 and 8:

$$\mathbb{A}_{i_{max}} \equiv FoV_{i_{min}} \tag{11}$$

In contrast, the $C^2_{n_i}$ value monotonically decreases as the value $\mathbb{A}_i$ is increased through Eq. 5. Therefore, the value of $C^2_{n_{i_{min}}}$ is obtained at $\mathbb{A}_{i_{max}}$. Consequently, $q = i$ is concluded, and from Eq. 11, we obtained $C^2_{n_{i_{min.}}} \equiv FOV_{i_{min}}$.

**Lemma 2**: Based on $C^2_{n_{i_{min}}} \equiv \mathbb{A}_{min}$ as expressed in Eq. 5, we conclude that $P_{M_{i_{min}}} \equiv C^2_{n_{i_{min}}}$. We obtained the following:

$$P_{M_{Min.}} \equiv FOV_{min.} \tag{12}$$

Lemma 2 indicates that when the $FOV$ of the digital drone camera has a minimum value, the power margin required for the optical channel is decreased, and the AT of such an optical channel is mitigated.

3. RESULTS

We employed MATLAB to investigate our results for air surveillance applications using the digital drone camera based FSO communication system performance. Table 2 lists the parameters used in the simulation. We explore the diverse $FOV$ parameters of the digital drone camera-based FSO system impression on the channel performance indicated by the $PM$, which is required to mitigate AT.

Table 2: Parameters used to obtain the results

| Parameter name | Range | Unit |
|---|---|---|
| Camera Pixel | 2000 | Pix. |
| Focal length | 10–180 | Mm |
| FOV | 5–1200 | Deg. |
| Image resolution($Ï$) | 100 | Pix/m |
| subarea HSL | 20 | M |
| $C_n^2$ terrestrial | 1 × 10-14 | m-2/3 |
| λ | 1550 | Nm |
| Optical Link (L) | 2, 5 | Km |

In Fig. 4, the selected digital drone camera pixel is 2000 Pix, $Ï$ is 100 Pix/m, and $L$ is 2 km, where the $FOV$ = 120–100 is stated. The optical channel performance is depicted by the outage probability ($Po$), which is measured in contrast to the required $PM$ owing to the AT effect on the optical channel. Our results reveal the diverse $FOV$ effects on the required $PM$ value. The $FOV$ values from maximum to minimum (that is, 120–100) permit an increase in the altitude of such a system (Eq. 8). To demonstrate the $FOV$ effects with a preselected $Ï$ value (i.e., result normalization), we assume that the horizontal-sidelong ($HSL$) of a scanned subarea by a single drone is constant (equal to 20 m), as illustrated in Fig. 1.The vertical-sidelong in the scanning direction and its length depend on the drone type and application (see Table 2).[2] Because $A$ is increased, the AT effects indicated by reducing the scintillation index $C_n^2$ are minimized using Eq. 5, accordingly. Thus, *PM* decreases correspondingly (Eq. 6) by reducing the *FOV* value of such a system, where $\mathbb{A}_{max}$ is obtained at $FOV_{min}$ (Eq. 8), and this result satisfies Lemma 1. As shown in Fig. 4, the *PM* required for a 2 km optical link length at Po = 10-10 is equal to 12.2 dB at *FOV* = 1200.  Meanwhile, for *FOV* = 900, the *PM* decreases from 12.2 to 11.2 dB, with a 1 dB gain in *PM* value of such a system. Subsequently, for a minimum value of *FOV* = 100, the required *PM* values are reduced and equal to approximately 2 dB only for such a system (that is, the *HSL* of a subarea that is scanned by a single drone and $\ddot{I}$ are constant and equal to 20 m and $\ddot{I}$ = 100 pix/m, respectively).

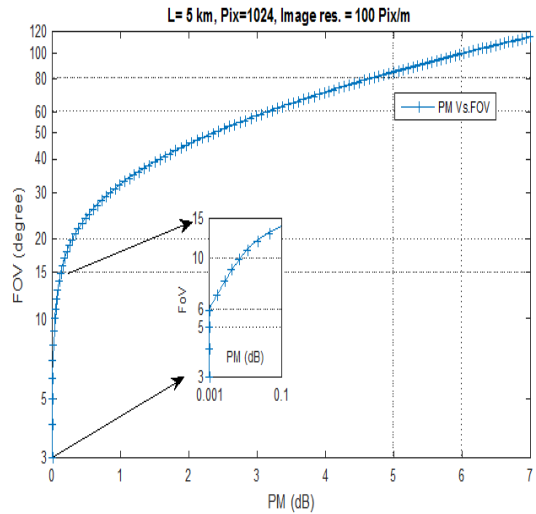

Fig. 4. Optical channel performances for different $FOV$ values at link length = 2 km.

the same camera pixels number and $\ddot{I}$, which are equal to 2000 Pix and 100 Pix/m, respectively. In this figure, the *PM* value required for $L = 5$ km is examined versus the *FOV*. As shown in Fig. 5, the required *PM* value decreased linearly with decreasing *FOV* values within the *FOV* range of
120–400. For *FOV* = 40–200, the required *PM* value declined exponentially. The decline in the required *PM* value became steep with values of FOV in the range of 20–60. For the value of *FOV* range of 6–30, the required *PM* value was approximately 0 dB. This occurred because the altitude was increased to a higher value (optimum), and the AT effect was mitigated. We assumed that the minimum value of the *FOV* was not less than 30 to ensure it was within the focal length range mentioned in Table 2. However, even if the imposed *FOV* is equal to 1, the *PM* value remains the same.

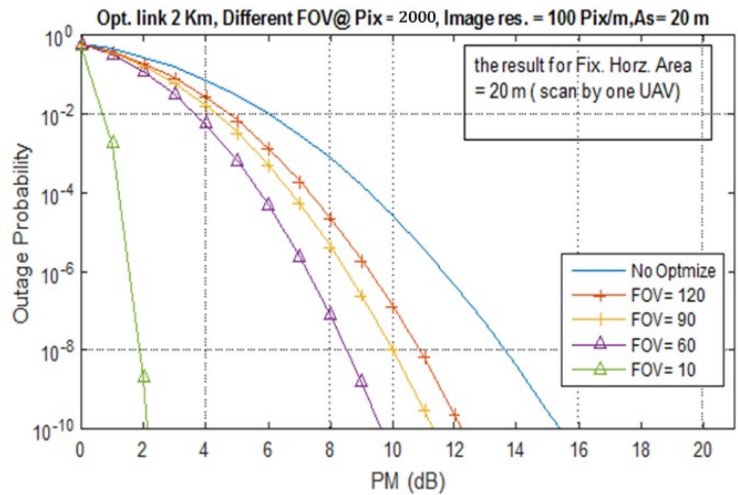

Fig. 5. Power margin required for different *FOVs* at 5 km optical link length

4. CONCLUSION

In this study, we combined a digital drone camera with optical channel parameters to increase the FSO system efficiency in air surveillance applications. We thereafter investigated the *FOV* of a digital drone camera with optical channel performance. The degradation of optical systems and optical channel altitude by AT are dependent on each other. This allows the two lemmas to couple the related parameters of both a digital drone camera and obtain an enhanced optical channel to mitigate the degrading effects of AT. The coupled lemmas led us to select the optimum *FOV* of a digital drone camera value out of the available range. This minimizes the AT effect and is indicated by the most critical parameter in Lemma 1. Consequently, Lemma 2 enhanced the (that is, reduced) value of the power margin required for the optical communication system through the preselected *FOV* of a digital drone camera. We quantified the performance improvements obtained for the optical link length schemes. For the preselected *FOV* values of 60 and 100 of a digital drone camera, the respective performance improvements increased up to 13 and 7 dB gain in the *PM* value, respectively. Compared to $FOV_{max}$, $FOV_{min}$ takes advantage of the altitude dependency of AT to a reduced degree and is outperformed by its counterpart as the *FOV* value decreases.